\documentclass[journal]{IEEEtran}
\ifCLASSINFOpdf
  \usepackage[pdftex]{graphicx}
\else
  \usepackage[dvips]{graphicx}
\fi
\newcommand{\reffig}[1]{Figure~\ref{#1}}

\usepackage[cmex10]{amsmath}
\usepackage{amssymb}
\usepackage{amstext}
\newtheorem{thm}{Theorem}[section]
\newtheorem{defn}[thm]{Definition}
\newtheorem{lem}[thm]{Lemma}
\newtheorem{prop}[thm]{Proposition}

\newcommand{\qeda}{\hfill $\Box$ \\}
\usepackage{array}
\begin{document}
%
\title{A Separation Algorithm for Improved \\LP-Decoding of Linear Block Codes}
%
%
%

\author{Akin~Tanatmis,~Stefan~Ruzika,~Horst~W.~Hamacher,\\
				Mayur~Punekar,~Frank~Kienle~and~Norbert~Wehn  
        
\thanks{A.~Tanatmis,~S.~Ruzika~and~H.~W.~Hamacher are with Department of Mathematics, University of Kaiserslautern, Erwin-Schroedinger-Strasse, 67663 Kaiserslautern, Germany. Email: \{tanatmis, ruzika, hamacher\}@mathematik.uni-kl.de}
\thanks{M.~Punekar,~F.~Kienle~and~N.~Wehn are with Microelectronic Systems Design Research Group, University of Kaiserslautern, Erwin-Schroedinger-Strasse, 67663 Kaiserslautern, Germany. Email: \{punekar, kienle, wehn\}@eit.uni-kl.de}

\thanks{This paper has been presented in part at the 5th International Symposium on Turbo Codes and Related Topics, September 1st - 5th, 2008, Lausanne, Switzerland.}
\thanks{Manuscript received December 02, 2008; revised;}}

%
%

\markboth{submitted to IEEE TRANSACTIONS ON INFORMATION THEORY}%
{Shell \MakeLowercase{\textit{et al.}}: Bare Demo of IEEEtran.cls for Journals}
%



\maketitle

\begin{abstract}
Maximum Likelihood (ML) decoding is the optimal decoding algorithm for arbitrary linear block codes and can be written as an Integer Programming (IP) problem. Feldman et al. relaxed this IP problem and presented Linear Programming (LP) based decoding algorithm for linear block codes. In this paper, we propose a new IP formulation of the ML decoding problem and solve the IP with generic methods. The formulation uses indicator variables to detect violated parity checks. We derive Gomory cuts from our formulation and use them in a separation algorithm to find ML codewords. We further propose an efficient method of finding cuts induced by redundant parity checks (RPC). Under certain circumstances we can guarantee that these RPC cuts are valid and cut off the fractional optimal solutions of LP decoding. We demonstrate on two LDPC codes and one BCH code that our separation algorithm performs significantly better than LP decoding. 
\end{abstract}

\begin{IEEEkeywords}
ML decoding, LP decoding, Integer programming, Separation algorithm.
\end{IEEEkeywords}

%
\IEEEpeerreviewmaketitle

\section{Introduction}
%
%
%
%
\IEEEPARstart{L}{ow-Density Parity-Check} (LDPC) codes have attracted significant interest in the research community in the last decade. LDPC codes are generally decoded by Belief Propagation (BP) (or Sum-Product) algorithm. BP exploits the sparse structure of the parity check matrix of LDPC codes very well and achieves good performance. However, due to the heuristic nature of BP algorithm, it is not possible to guarantee the performance of BP decoders at very low error rates. Moreover, the performance of BP is very poor for arbitrary linear block codes with dense parity check matrices (which means that the corresponding Tanner graph contains short cycles).

ML decoding of linear block codes can be modeled as an IP problem. However, since the ML decoding is NP-hard \cite{BeMcTi}, solving this IP problem is computationally feasible only for small instances. Nevertheless considering ML decoding as an IP problem yields a new approach to derive sub-optimal algorithms. These algorithms offer some advantages compared to BP decoding. First, these approaches rely on a well-studied mathematical theory which enables quantitative statements (e.g. convergence, complexity, correctness, etc.) with regard to the decoding process and its result \cite{DiGoWa}, \cite{FeWaKa}, \cite{TagSie}. Secondly, they are not limited to sparse matrices.

In \cite{FeWaKa} Feldman et al. proposed a new algorithm based on LP to decode binary linear codes. This LP decoding algorithm utilizes a set of constraints which contains all valid codewords of a given code and a linear objective function. Minimizing this objective function over the resulting polytope yields the ML codeword if the optimal solution is integral (known as ML certificate property \cite{FeWaKa}). If the optimal solution is not integral then LP decoder outputs an error.

Recently, LP decoding has been improved towards lower complexity (\cite{Burshtein}, \cite{ChertStep}, \cite{TagSie}, \cite{VoKo}, \cite{YaFeWa}, \cite{YaWaFe} ) and better perfomance (\cite{ChertChern}, \cite{chertkov-2007}, \cite{DiGoWa}, \cite{DrYeYi}). Analysis of error correction performance of LP decoding (\cite{DaDiKaWa}, \cite{FeMa}, \cite{VoMOLB}) and the relationship to iterative message passing algorithms (\cite{FeWaKa}, \cite{VoKoGraph}, \cite{VoKoe}) have also been studied in the literature. 

In this paper, we concentrate on improving linear programming decoding using a separation algorithm. We introduce an alternative IP formulation for the decoding problem. Instead of solving the optimization problem, we attempt to find the ML solution by an iterative separation approach: First, we relax the IP formulation and solve the resulting linear program. In case of a non-integral optimal solution, we derive inequalities which cut off this non-integral solution, add these inequalities to the LP formulation and resolve the LP problem. This process continues until an optimal integer solution is found or further cuts cannot be generated. It should be noted that this general integer programming approach known as separation problem has first been applied to LP decoding by Taghavi and Siegel \cite{TagSie}. Our approach offers however the following advantages which remarkably facilitate LP based decoding. 
\begin{enumerate}
\item The number of constraints in the new IP formulation is the same as the number of rows in the parity check matrix. Each parity check equation which is originally in $GF(2)$ is converted into a linear constraint in $\mathbb{R}^n$ by means of an auxiliary variable.  
\item The auxiliary variables serve as indicators which can be used for identifying violated parity check constraints. We can prove that we detect violated inequalities faster than the adaptive algorithm of Taghavi and Siegel under some mild assumptions. 
\item We formally show that the Forbidden Set Inequalities \cite{DiGoWa} are a subset of the set of Gomory cuts (see \cite{NemWol}) which can be deduced from our formulation.
\item We provide empirical evidence that our new separation algorithm performs better than LP decoding. This is mainly due to generating strong cuts efficiently using alternative representations of the codes at hand. 
\end{enumerate}
To provide empirical evidence we applied the \scshape New Separation Algorithm \normalfont to decode two LDPC codes along with one BCH code. 

The rest of this paper is organized as follows. We introduce notation in Section II and briefly review relevant literature in Section III. In Section IV, we introduce the new IP formulation, its LP relaxation, and the \scshape New Separation Algorithm \normalfont. In Section V we present our numerical results and compare them with BP, LP decoding, and the lower bound resulting from ML decoding. The paper is concluded with some remarks and further research ideas in Section VI. 
\section{Notation and background}
A binary linear block code with cardinality $2^k$ and block length $n$ is a $k$ dimensional subspace of the vector space $\{0,1\}^n$ defined over the field $GF(2)$. The linear code $C$ is given by $k$ basis vectors of length $n$ which are represented by a $k \times n$ matrix $G$ (generator matrix). Equivalently $C$ can be described by a parity check matrix $H \in \{ 0,1 \}^{m \times n}$ where $m=n-k$.We thus have $x \in C$, i.e. $x$ is a codeword, if and only if $Hx=0$ in $GF(2)$. We denote the $i^{th}$ row and $j^{th}$ column of $H$ by $H_{i,.}$, $H_{.,j}$ respectively. $H_{i,.}x=0$ in $GF(2)$ is defined as the $i^{th}$ parity check constraint. The index set $I=\left\{ 1, \ldots, m\right\}$ refer to the rows and the index set $J=\left\{ 1, \ldots, n\right\}$ refer to the columns of $H$. The matrix $H$ is often represented by a Tanner graph $\mathbb{G}=(V,E)$. The node set $V$ of $\mathbb{G}$ consists of the two disjoint node sets indexed by $I$ and $J$ called the check nodes and variable nodes respectively. An edge $[i,j] \in E$ connects node $i$ and $j$ if and only if $H_{ij}=1$.

The ML decoding problem for any binary code $C \in \{0,1\}^n$ can be written in terms of the mathematical program  
\begin{eqnarray}
\min \{ c^Tx :x \in C \}&=& \min \{ c^Tx :x \in \text{conv}(C) \} \label{eq:MLdecoding}.
\end{eqnarray} 

\noindent{Here, $c \in \mathbb{R}^n$ is the cost vector obtained by the log-likelihood ratios $c_i = \text{log} \left( \frac{P(\hat{x}_i|x_i=0)}{P(\hat{x}_i|x_i=1)} \right)$ for a given received bit $\hat{x}_i$ and $\text{conv}(C)$ denotes the convex hull of $C$ i.e. the codeword polytope. The left hand side of the equation (\ref{eq:MLdecoding}) is an integer programming problem which is known to be NP-hard \cite{BeMcTi}. Replacing $C$ with $\text{conv}(C)$ leads to a linear programming problem which is stated on the right hand side of (\ref{eq:MLdecoding}). Although linear programming is polynomially solvable in general, computing $\text{conv}(C)$ is intractable. In other words a concise description of $\text{conv}(C)$ by means of linear inequalities increases exponentially in the block length $n$. Thus ML decoding remains a challenging task. Nevertheless, linear programming decoding can be applied efficiently if good approximations of the codeword polytope can be found. Recently attempts in this direction have been made, (e.g.\cite{ChertStep}, \cite{FeWaKa}, \cite{TagSie}, \cite{VoKo}, \cite{YaWaFe}).

Feldman et al. \cite{FeWaKa} introduced the LP decoder which minimizes $c^Tx$ over a relaxation of the codeword polytope. The relaxation is achieved by using the parity check matrix $H$. Each row (check node) $i \in I$ defines a local code $C_i$, i.e. local codewords $x \in C_i $ are the bit sequences which satisfy the $i^{th}$ parity check constraint. Note that $C=C_1 \cap \ldots \cap C_m$. 
\begin{lem}[\cite{VoKo}]  \label{lem:lemfundpoly}
Let $P=\text{conv}(C_1)  \cap  \ldots \cap \text{conv}(C_m)$. If $C=C_1  \cap  \ldots  \cap C_m$ then $\text{conv}(C) \subseteq P $.  
\end{lem}

$P$ is generally referred to as the fundamental polytope (\cite{DiGoWa}, \cite{TagSie}, \cite{VoKoGraph}). This relaxation has the advantage that the complexity of describing the convex hull of any local code $\text{conv}(C_i)$ and thus of $P$ is much less than the complexity of describing the codeword polytope $C$. The LP decoder solves the problem $\min \{ c^Tx :x \in P \}$. 

Several approaches are used in \cite{ChertStep}, \cite{FeWaKa}, \cite{TagSie}, \cite{VoKo} \cite{YaWaFe} to write constraints completely describing $P$. 
We are going to use the set of constraints already introduced in \cite{FeWaKa} and referred to as Forbidden Set Inequalities in \cite{DiGoWa}. The index set of variable nodes which are adjacent to check node $i$ is defined as $N_i:=\left\{j \in J: H_{ij}=1 \right\}$. Using $S \subseteq N_i$ we assign values to code bits $x_j$ as follows. Set $x_j=1$ for all $j \in S$, and $x_j=0$ for all $ j \in N_i \setminus S$. For $j \notin N_i$, $x_j$ can be chosen arbitrarily. 
These value assignments to variables are feasible, i.e. satisfy the parity check constraint, for the local code $C_i$ if $\left|S\right|$ is even. If $\left|S\right|$ is odd, they are, however, infeasible or forbidden. From this observation the so called Forbidden Set Inequalities are derived. Let $\Sigma_i=\left\{S \subseteq N_i : \left| S \right| \; \textnormal{odd} \right\}$. It is shown in \cite{FeWaKa} that $\text{conv}(C_i)$ can be described by 

\begin{eqnarray}
&& \sum_{j \in N_i \setminus S} x_j + \sum_{j \in S} (1- x_j) \geq 1 \; \forall S \in \Sigma_i \label{forbidset}
\end{eqnarray}

which can equivalently be written as 

\begin{eqnarray}
&& \sum_{j \in S} x_j - \sum_{j \in N_i \setminus S} x_j \leq \left| S \right| - 1 \; \forall S \in \Sigma_i. \label{forbidsetnew}
\end{eqnarray}

Consequently the LP decoder solves
\begin{align*}
&  \min c^Tx \qquad \qquad \qquad \qquad \qquad \qquad \qquad \qquad \text{(LPD)} \\ 
& \text{s.t.} \sum_{j \in S} x_j - \sum_{j \in N_i \setminus S} x_j  \leq  \left| S \right| - 1 \; \forall  S \in \Sigma_i , \; i=1, \ldots ,m \\
& 0  \leq x  \leq 1. \\
 \nonumber
\end{align*}
If LPD has an integral optimal solution then the LP decoder outputs the ML codeword. If LPD has a non-integral optimal solution then the LP decoder outputs an error. The number of Forbidden Set Inequalities induced by check node $i$ is $2^{\delta(i)-1}$ where $\delta(i)=\sum_{j=1}^n H_{ij}$ is the check node degree, i.e. the number of edges incident to node $i$. The LP decoder can thus be applied successfully to low density codes. As the check node degrees increase the computational load of building and solving the LP model is however in general prohibitively large. This makes the explicit description of the fundamental polytope via Forbidden Set Inequalities inapplicable for high density codes. To overcome this difficulty an alternative formulation which requires $O(n^3)$ constraints is proposed in \cite{FeWaKa}. More recent formulations of \cite{ChertStep} and \cite{YaWaFe} have size linear in the length and check node degrees. 
Another approach applicable to high density codes is to solve the corresponding separation problem of LPD \cite{TagSie}. The separation problem over an implicitly given polyhedron is defined as follows: 
\begin{defn} \label{defn:SepAlg}
Given a bounded rational polyhedron $P \subset \mathbb{R}^n$ and a rational vector $x^* \in \mathbb{R}^n$, either conclude that $x^* \in P$
or, if not, find a rational vector $(\Pi,\Pi_0) \in \mathbb{R}^n \times \mathbb{R}$ such that $\Pi^Tx \leq \Pi_{0}$ and $\Pi^Tx < \Pi^Tx^*$ for all $x \in P$. In the latter case $(\Pi,\Pi_0)$ is called a valid cut.
\end{defn}
In separation algorithms (see \cite{NemWol}) one iteratively computes families $\Lambda$ of valid cuts until no further cuts can be found. In the separation algorithm of \cite{TagSie}, which is called adaptive LP decoding by the authors, Forbidden Set Inequalities are not added all at once in the beginning as in \cite{FeWaKa} but iteratively. In other words, the separation problem for the fundamental polytope is solved by searching violated Forbidden Set Inequalities. In the initialization step of the LP $\min \{ c^Tx : 0 \leq x \leq 1 \}$ is computed. An optimal solution $x^*$ is checked in $O(m\delta^{max}+n\text{log}n)$ time, if $x^*$ violates any forbidden set inequality where $\delta^{max}$ is the maximum check node degree. If some of the Forbidden Set Inequalities are violated then these inequalities are added to the formulation and the LP is resolved including the new inequalities. 

Adaptive LP decoding stops when the current optimal solution $x^*$ satisfies all Forbidden Set Inequalities. If $x^*$ is integral then it is the ML codeword otherwise an error is output. Note that putting the LP decoder in an adaptive setting does not yield an improvement in terms of frame error rate since the same solutions are found. On the other hand the adaptive LP decoder converges with less constraints than the LP decoder which has a positive effect on computation time.

The communication performance of LP decoding motivated researchers to find better approximations of the codeword polytope as part of ML decoding. One way is to tighten the fundamental polytope with new valid inequalities. Among some other generic techniques of cut generation, adding so called RPC cuts is proposed in \cite{FeWaKa}. Redundant parity checks are obtained by adding a subset of rows of $H$ matrix in $GF(2)$. These checks are redundant in the sense that they do not alter the code (they may even degrade the performance of BP \cite{FeWaKa}). However they induce new constraints in the LP formulation which may cut off a particular non-integral optimal solution thus tightening the fundamental polytope. An open problem is to find methods to generate redundant parity checks efficiently such that the induced constraints are guaranteed to cut off a non-integral LP solution. 

To the best of our knowledge two approaches for generating potential cuts exist so far. First, adding redundant parity check cuts which result from adding any two rows of $H$ \cite{FeWaKa}. Secondly, the approach in \cite{TagSie} which makes use of the cycles in the Tanner graph: 1) given a non-integral optimal solution $x^*$ remove all variable nodes $j$ form the Tanner graph for which $x_j^*$ is integral; 2) find a cycle by randomly walking through the pruned Tanner graph; 3) add the rows of the $H$ matrix in $GF(2)$ which correspond to the check nodes in the cycle; 4) check if the found RPC introduces a cut. 
\section{A New Separation Algorithm Based on an alternative IP Formulation}
Our separation algorithm is based on the following formulation which we refer to as Integer Programming Decoding $(IPD)$. 
\begin{align*}
& \min  c^Tx  \qquad \qquad  \text{(IPD)} \\
\text{s.t.} \ & Hx - 2 z = 0 \\
&x \in \left\{ 0,1 \right\}^n\\
& z \geq 0, \; \textnormal{integer}
\end{align*}
IPD is an integer programming problem which works as an ML decoder. The auxiliary variable $z \in \mathbb{Z}^m$ ensures the binary constraint $Hx=0$ over $GF(2)$ turns into a constraint over the real number field $\mathbb{R}$ which is much easier to handle. This formulation has the additional advantage that the number of constraints is the same as the number of rows of the parity check matrix. Note that LPD can also be used as an ML decoder by restricting $x$ to be in $\{0,1\}^n$. Yet in this case the number of constraints is exponential in the check node degree. Although our formulation IPD has less constraints, this does not change the fact that ML decoding is NP-hard. Therefore our approach is to solve the separation problem by iteratively adding new cuts $\Pi^Tx \leq \Pi_{0}$ according to Definition \ref{defn:SepAlg} and solving the LP relaxation of IPD given by 
\begin{align*}
&\min  c^Tx \qquad \qquad \text{(RIPD)}\\
\text{s.t.} \ & Hx - 2 z = 0 \\
&\Pi^Tx \leq \Pi_{0} \quad (\Pi,\Pi_0) \in \Lambda\\
&0 \leq x \leq 1\\
& z \geq 0. 
\end{align*}
 
Note that in the initialization step there are no cuts of type $\Pi^Tx \leq \Pi_{0}$ i.e. $\Lambda=\emptyset$. If RIPD has an integral solution $(x^*,z^*) \in \mathbb{Z}^{n+m}$ then $x^*$ is the ML codeword. Otherwise we generate cuts of the type $\Pi^Tx \leq \Pi_{0}$ in order to exclude the non-integral solution found in the current iteration. We add these inequalities to the formulation and solve RIPD again. In a non-integral solution of RIPD $x$ or $z$ (or both) is non-integral. If $x \in \mathbb{Z}^n$ and $z \in \mathbb{R}^m \setminus \mathbb{Z}^m$ then we add Gomory cuts (see \cite{NemWol}) which is a generic cut generation technique used in integer programming. Surprisingly, in this case Gomory cuts can be shown to correspond to Forbidden Set Inequalities.      

\begin{thm} \label{thm:theorem1}
Let $(x^*,z^*) \in \mathbb{Z}^n \times \mathbb{R}^m$ be the optimal solution of RIPD such that $z^*_i \in \mathbb{R} \setminus \mathbb{Z}$  for $i \in I$. Then the Gomory cut which is violated by $(x^*,z^*)$ is the Forbidden Set Inequality 
\begin{align}
\sum_{j \in S} x_j - \sum_{j \in N_i \setminus S} x_j \leq \left| S \right| - 1 \label{eq:Gomeq}
\end{align}
where $S := \left\{ j \in N_i \; | \; x^*_j=1 \right\}$.
\end{thm}

\noindent{\textbf{Proof:}}\\
We apply the general method known as Gomory's cutting plane algorithm (see e.g. \cite{NemWol}) to our special case. Gomory cuts are derived from the rows of the simplex tableau in order to cut off non-integral LP solutions and find the optimal solution to the integer linear programming problems. 
Consider RIPD at any step: 
\begin{align*}
&\min  c^Tx \qquad \qquad \text{(RIPD)}\\
\text{s.t.} \ &Hx - 2 z = 0 \\
&0 \leq x \leq 1\\
&Ax \leq b\\
& z \geq 0 
\end{align*}
\noindent{$\text{where } c,x \in \mathbb{R}^n, \; H \in \{ 0,1 \}^{m \times n}, \; z  \in \mathbb{R}^m, \; A \in \{ -1,0,1 \}^{\lambda \times n}$ for some $\lambda \in \mathbb{N}_0$ and $b \in \mathbb{N}_0^\lambda$. Note that $\lambda$ is the number of constraints added iteratively until the current step, i.e. $\lambda=\left| \Lambda \right|$. The $\lambda \times n$ matrix $A$ is the coefficient matrix of the iteratively added constraints, i.e. $\Pi^Tx \leq \Pi_{0} \quad (\Pi,\Pi_0) \in \Lambda$. We denote the right hand sides of these constraints with the vector $b$. RIPD in standard form can be written as follows:
\begin{align}
&\min  c^Tx \qquad \qquad \text{(RIPD)} \label{lp1}\\
\text{s.t.} \ &z - \bar{H}x = 0 \label{lp2} \\
& x + s_1 = 1 \label{lp3}\\
&A x + s_2 = b \label{lp4}\\
& z \geq 0, \; x \geq 0, \; s \geq 0. \label{lp5} 
\end{align}
\noindent{where $\bar{H}:=\frac{1}{2}H$, $s=(s_1,s_2) \in \mathbb{R}^{n+\lambda}$.}  
For ease of notation we rewrite (\ref{lp1})-(\ref{lp5}) as 
\begin{align}
& \min \bar{c}^T y \label{standardform1} \\
\text{s.t.} \ & Py=q \label{standardform2} \\
& y \geq 0. \label{standardform3}
\end{align}
Note that
\begin{align*}
\bar{c}^T&=(\bar{c}_1, \ldots, \bar{c}_m, \bar{c}_{m+1}, \ldots, \bar{c}_{m+n}, \bar{c}_{m+n+1}, \ldots, \bar{c}_{m+2n+\lambda})\\
&=(0, \ldots, 0, c_1, \ldots, c_n, 0, \ldots, 0),\\
y^T&=(y_1, \ldots, y_m, y_{m+1}, \ldots, y_{m+n}, y_{m+n+1}, \ldots, y_{m+2n+\lambda})\\
&=(z_1, \ldots, z_m, x_1, \ldots, x_n, s_1, \ldots, s_{n+\lambda}) \text{ and}\\
q^T&=(q_1, \ldots, q_m, q_{m+1}, \ldots, q_{m+n}, q_{m+n+1}, \ldots, q_{m+2n+\lambda})\\
&=(0, \ldots, 0, 1, \ldots, 1, b_1, \ldots, b_{\lambda}).
\end{align*}
The constraint matrix $P$ has $m+n+\lambda$ rows and $m+2n+\lambda$ columns. We denote the $\alpha^{th}$ row of $P$ with $P_\alpha$ where $\alpha \in \{ 1,\ldots,m+n+\lambda \}$ and $\beta^{th}$ column of $P$ with $P^\beta$ where $\beta \in \{ 1,\ldots,m+2n+\lambda \}$. The component in row $\alpha$ and column $\beta$ is denoted with $P_{\alpha \beta}$. Additionally, we define the $\alpha^{th}$ unit vector as $e^\alpha \in \mathbb{R}^{m+n+\lambda}$. Thus, we rewrite $P$ as
\begin{align*}
&P=\left[e^1 \ldots e^m P^{m+1} \ldots P^{m+n} e^{m+n+1} \ldots e^{m+2n+\lambda} \right].
\end{align*}
The first $m$ columns of the constraint matrix $P$ are the unit vectors corresponding to the variables $\{z_1 \ldots z_m\}$. Likewise, the last $n+\lambda$ columns are the unit vectors corresponding to the slack variables $\{s_1 \ldots s_{n + \lambda} \}$.   

The first $m$ linear equations of $Py=q$ are of the form: 
\begin{align*}
z_i -\frac{1}{2}\cdot\sum_{j \in N_i}x_j=0 \; \text{for all } i \in \{1, \ldots, m \}.
\end{align*}
Let $y^*=(z^*,x^*,s^*) \in \mathbb{R}^{m+2n+\lambda}$ be the optimal solution to (\ref{lp1})-(\ref{lp5}). By assumption it is $x^* \in \{ 0,1 \}^n$. For $i \in \{1, \ldots, m \}$, $z_i$ is given by $z^*_i=\frac{1}{2}k_i$, where       
\begin{displaymath}
k_i=\left|\{ j \in N_i | x^*_j=1 \}\right|.
\end{displaymath} 
It is obvious that $k_i \in \mathbb{N}_0$. If $k_i$ is even i.e. an even number of variable nodes are set to $1$ in the neighborhood of the check node $i$, then $z^*_i \in \mathbb{N}_0$ holds. Otherwise, $z_i$ is an odd multiple of $\frac{1}{2}$. We then consider the Gomory cut for this row $i$.

For the optimal solution $y^*$ we can partition $P$ into a basis submatrix $P_B$ and a non-basis submatrix $P_N$, i.e. $P=\left[P_B \; P_N\right]$. Let $B$ and $N$ denote the index sets of the columns of $P$ belonging to $P_B$ and $P_N$, respectively. An  $(m+n+\lambda) \times (m+n+\lambda)$ basis matrix, $P_B$, corresponding to the optimal solution $y^*$ can be constructed as follows. First we take the columns $e^1,\ldots,e^m$ which are the identity vectors corresponding to the variables $\{z_1 \ldots z_m\}$ into $P_B$. Secondly for $j=1,\ldots,n$, we include the column $P^{m+j}$ if $x^*_j=1$ or $P^{m+n+j}$ if $s^*_j=1$ in $P_B$. There exists $n$ such columns since
\begin{align*}
\sum_{j=1}^n (x^*_j+s^*_j)=n  
\end{align*}  
must hold due to (\ref{lp3}). Finally we take the columns $e^{m+2n+1}, \ldots, e^{m+2n+\lambda}$ corresponding to the slack variables which are written for the iteratively added constraints. The variables corresponding to the columns in the basis matrix are called basic variables. The remaining columns of $P$ form the non-basis submatrix $P_N$. The columns of $P_N$ are the columns $P^{m+j}$, $j=1,\ldots,n$, for which $x^*_j=0$ and the columns $e^{m+n+j}$ ,$j=1,\ldots,n$, for which $s^*_j=0$. The variables corresponding to the columns in $P_N$ are called non-basic variables. 

The Gomory cut for row $i$ of $P$ is given by the inequality 
\begin{eqnarray}
&& \sum_{h \in N} \left( \bar{p}_{ih} -  \left\lfloor  \bar{p}_{ih} \right\rfloor  \right)y_h \geq \left( \bar{q}_i -  \left\lfloor \bar{q}_i \right\rfloor \right)  
\end{eqnarray}
where $\bar{p}_{ih}=(P_B^{-1})_i\cdot(P_N)^h$, and $\bar{q_i}=(P_B^{-1})_i \cdot q$. Note that in our case $i \leq m$ since only $z^*$ has non-integral components. In the following we investigate the structure of $(P_B^{-1})_i$, $(P_N)^h$, $\bar{p}_{ih}$ and $\bar{q}_i$. 

For a fixed $i$, it can easily be verified that the entries $(P_B^{-1})_{il}$, $l=1,\ldots,m+n+\lambda$ of $(P_B^{-1})_i$ are given as
\begin{eqnarray*}
(P_B^{-1})_{il} &=&\left\{
\begin{array}{rl}
1, & \text{if } l=i\\
\frac{1}{2}, & \text{if } P_{il}=1, x^*_j = 1, \\
     & l=m+j, j=1, \ldots, n \\
0 & \text{otherwise} 
\end{array} \right.
\end{eqnarray*} 
\noindent{(This can be verified by observing the changes on row $i$ when we append an $(m+n+\lambda) \times (m+n+\lambda)$ identity matrix to $P_B$ and perform the Gauss-Jordan elimination on the appended matrix in order to get $P_B^{-1}$.)}

Having found $(P_B^{-1})_i$, $\bar{q}_i$ is then computed by
\begin{align}
\bar{q}_i&=(P_B^{-1})_i\cdot q \\
& =q_i+\frac{1}{2}\sum_{j:x^*_j=1}q_{m+j}\\
& =0+\frac{1}{2}\sum_{j:x^*_j=1}1.
\end{align}
Thus, we showed that $\bar{q}_i$ is $\frac{1}{2}$ times the number of basic $x$ variables in row $i$. Since $z_i$ is not integer, the number of basic $x$ variables in row $i$ is odd. It follows that in our case the right hand side of the Gomory cut, $\bar{q}_i - \left\lfloor \bar{q}_i \right\rfloor$, is always $\frac{1}{2}$.

Next, we compute $\bar{p}_{ih}=(P_B^{-1})_i \cdot (P_N)^h$. The columns of $P_N$ are the columns of $P$ corresponding to non-basic $x$ components (i.e. $x^*_j=0$) and non-basic $s$ components (i.e. $s^*_j=0$)  $j=1,\ldots,n$. If
$(P_N)^h=P^{m+j}$ such that $x^*_j=0$, then for a fixed value of $h$, the entries of $(P_N)^h$, $(P_N)_{oh}$, $o=1,\ldots,m+n+\lambda$ are given as 
\begin{eqnarray*}
(P_N)_{oh} &=&\left\{
\begin{array}{rl}
-\frac{1}{2}, & \text{if } P_{o(m+j)}=1 \text{ and } o \leq m \\
1, & \text{if } o=m+j\\
0 & \text{otherwise.} 
\end{array} \right.
\end{eqnarray*}  
If $(P_N)^h=P^{m+j}$ such that $s^*_j=0$, then $(P_N)^h$ is the unit vector $e^{m+j}$. 

For the case that $(P_N)^h=P^{m+j}$ where $x^*_j=0$, the only position where both $(P_B^{-1})_i$ and $(P_N)^h$ may have nonzero entries is position $i$. For all other positions $l=1,\ldots,m+n+\lambda$ and $l \neq j$ either $(P_B^{-1})_{il}=0$ or $(P_N)_{lh}=0$. This implies

\begin{eqnarray*}
\bar{p}_{ih}=(P_B^{-1})_i(P_N)^h &=&\left\{
\begin{array}{rl}
-\frac{1}{2}, & \text{if } P_{ih}=1  \\
0, & \text{if } P_{ih}=0. \\
\end{array} \right.
\end{eqnarray*} 
   
For the case that $(P_N)^h=P^{m+j}$ where $s^*_j=0$, position $m+j$ is the only position where both $(P_B^{-1})_i$ and $(P_N)^h$ may have a nonzero entry. This means, $\bar{p}_{ih}=(P_B^{-1})_i(P_N)^h=\frac{1}{2}$ for all non-basic $s$ variables corresponding to the basic $x$ variables in row $i$. If we denote the non-basic $x$ variables in row $i$ with the index set $N_i \setminus S:=\{j:x^*_j=0\}$ and the non-basic $s$ variables corresponding to the basic $x$ variables in row $i$ with the index set $S:=\{j:s^*_j=0\}$, we can write the Gomory cut as
\begin{eqnarray}
&&\sum_{h \in N} \left( \bar{p}_{ih} -  \left\lfloor  \bar{p}_{ih} \right\rfloor  \right)y_h \geq \frac{1}{2} \nonumber\\
&& \Leftrightarrow \sum_{j \in N_i \setminus S} \left( -\frac{1}{2} -  \left\lfloor  -\frac{1}{2} \right\rfloor  \right)x_j + \sum_{j \in S} \left( \frac{1}{2} -  \left\lfloor  \frac{1}{2} \right\rfloor  \right)s_j \geq \frac{1}{2} \nonumber \\
&& \Leftrightarrow \sum_{j \in N_i \setminus S} \frac{1}{2} x_j + \sum_{j \in S} \frac{1}{2}s_j \geq \frac{1}{2} \nonumber \\
&& \Leftrightarrow \sum_{j \in N_i \setminus S} x_j + \sum_{j \in S}(1-x_j) \geq 1. \label{gomcut}
\end{eqnarray}
Since inequality (\ref{gomcut}) is the forbidden set inequality obtained from the configuration $S := \left\{ j \in N_i \; | \; x^*_j=1 \right\}$ this concludes the proof. \qeda 

Given an optimal solution of RIPD, $(x^*,z^*)$ with $x^*_j \in \{ 0,1 \}$ for all $j \in J$ and $z^*_i \in \mathbb{R} \setminus \mathbb{Z}$ for at least one $i \in I$ we can efficiently derive Gomory cuts with the following algorithm.
\begin{tabular}{|p{0.9\columnwidth}|} 
\hline \\
\textbf {\underline{\scshape Cut Generation Algorithm \normalfont $1$}}\\\\
\textbf {Input} :  $(x^*,z^*)$ such that $x^*$ integral, $z^*$ non-integral. \\
\textbf {Output} : Gomory cut(s).\\
\textbf {1} : Set $i=1$. \\
\textbf {2} : If $k_i=2z^*_i$ is odd go to 3. Otherwise go to 5. \\
\textbf {3} : Set configuration $S := \left\{ j \in N_i \; | \; x^*_j=1 \right\}$.  \\
\textbf {4} : Construct constraint (\ref{eq:Gomeq}). \\
\textbf {5} : If $i \leq m$, set $i = i+1$ go to 2. Otherwise terminate. \\
\hline
\end{tabular}\\[0.1cm]
This algorithm has a computational complexity of $O(m\delta^{max})$ because at most $m$ values have to be checked until a violated parity check constraint is identified and $O(\delta^{max})$ is the complexity of constructing (\ref{eq:Gomeq}). An algorithm to check if any forbidden set inequality is violated is also given in \cite{TagSie}. In order to find a violated forbidden set inequality, the algorithm of Taghavi and Siegel first sorts $x$. Next, at most $\delta^{max}$ Forbidden Set Inequalities have to be generated and validated. Repeating this procedure for $m$ check nodes leads to an algorithm of time complexity $O(m\delta^{max}+ n\text{log}n)$. In contrast, we can efficiently determine the violated parity checks using the indicator variables $z$. Having identified a violated parity check constraint $i$ (if there exists any) we construct $(\ref{eq:Gomeq})$ easily by setting the coefficient of $x_j$ for $\{j \in N_i:x^*_j=1 \}$ to $+1$, the coefficient of $x_j$ for $\{j \in N_i:x^*_j=0 \}$ to $-1$ and $\left| S \right|=k_i$.  

Next we consider the situation that $0<x_j^*<1$ for some $j \in J$. Although it is still possible to derive a Gomory cut, \scshape Cut Generation Algorithm \normalfont $1$ is not applicable since Theorem \ref{thm:theorem1} holds only for integral $x^*$. For non-integral $x^*$ we propose the following separation method in order to find valid cutting inequalities, the \scshape Cut Generation Algorithm \normalfont $2$. The idea behind \scshape Cut Generation Algorithm \normalfont $2$ is based on Proposition \ref{prop1} and Proposition \ref{prop2}. 
\begin{prop} \label{prop1}
The Forbidden Set Inequalities derived from row $i$, $i \in \{ 1, \ldots,  m \}$, of a parity check matrix $H$ and the inequalities $0 \leq x \leq 1$, completely describe the convex hull $\text{conv}(C_i)$ of the local codeword polytope $C_i$. 
\end{prop}
\noindent{\textbf{Proof:}} This is shown in Theorem 4 in \cite{FeWaKa}. \qeda
\begin{prop} \label{prop2}
Let $x^*$ be a non-integral optimal solution of RIPD and $x^* \in$ conv$(C_i)$. Then there are at least two indices $j,k \in J$ such that $0<x_j<1$ and $0<x_k<1$. In other words check node $i$ cannot be adjacent to only one non-integral valued variable node.  
\end{prop}
\noindent{\textbf{Proof:}} If $x^* \in \text{conv}(C_i)$ then it can be written as a convex combination of two or more extreme points of conv$(C_i)$. Next we make use of an observation given in the proof of Proposition $1$ in \cite{DiGoWa}. Assume that check node $i$ is adjacent to only one non-integral variable node. This implies that there are two or more extreme points of conv$(C_i)$ which differ in only one bit. Extreme points of conv$(C_i)$ differ however, in at least two bits since they all satisfy parity check $i$ which contradicts the assumption. \qeda  

A given binary linear code $C$ can be represented with some alternative, equivalent parity check matrix which we denote with $\hat{H}$. Any such alternative parity check matrix for $C$ is obtained by performing elementary row operations on $H$. Note that Proposition \ref{prop1} is valid for any $\hat{H}$. Likewise Proposition \ref{prop2} holds as well for the parity check nodes $i \in \{1, \ldots, m\}$ of the Tanner graph representing $\hat{H}$. The rows of $\hat{H}$ may also be interpreted as redundant parity checks. Given a non-integral optimum $x^*$ of RIPD, in \scshape Cut Generation Algorithm \normalfont $2$ we search for a parity check which is adjacent to only one non-integral valued variable node. If we find such a parity check we know due to Proposition \ref{prop2} that $x^*$ can not be in the convex hull of this particular parity check. Furthermore due to Proposition \ref{prop1} there exists a forbidden set inequality which cuts off $x^*$. Note that in an exhaustive search algorithm one would check $2^m$ redundant parity checks if the parity check is adjacent to only one non-integral valued variable node. 

Instead of a computationally expensive exhaustive search we propose the \scshape Construct $\hat{H}$ Algorithm \normalfont which resembles Gaussian elimination. We transfer matrix $H$ into an equivalent matrix $\hat{H}$ by elementary row operations (adding two rows is in $GF(2)$). Our aim is to represent code $C$ with an alternative parity check matrix $\hat{H}$, so that in row $\hat{H}_{i,.}$ there exists exactly one $j \in J$ where $\hat{H}_{i,j}=1$ and $x^*_j$ is non-integral. For all other indices $h \in J\setminus \{j\}$ with $\hat{H}_{i,h}=1$, $x^*_h$ is integral. The \scshape Construct $\hat{H}$ Algorithm\normalfont tries to convert columns $j$ of $H$ with $x^*_j \notin \mathbb{Z}$ into unit vectors. Note that at most $m$ columns of $H$ are converted.\vspace{1mm}
\begin{tabular}{|p{0.9\columnwidth}|} 
\hline \\
\textbf {\underline{\scshape Construct $\hat{H}$ Algorithm \normalfont }}\\\\
\textbf {Input} :  $(x^*,z^*)$ such that $x^*$ non-integral \\
\textbf {Output} : $\hat{H}$.\\
\textbf {1} : Set $l=1$, $j=1$. \\
\textbf {2} : If $x_j^* \in (0,1)$ then go to 3. Else go to 4.\\
\textbf {3} : If $l \leq m$ then do elementary row operations until $H_{l,j}=1$ and $H_{i,j}=0$ for all $i \in I \setminus \{l\}$. Set $l=l+1$.\\
\textbf {4} : Set $j=j+1$. If $j \leq n$ then go to 2. Otherwise terminate.\\
\hline
\end{tabular}\\[0.1cm]
$\hat{H}$ can be obtained in $O(m^2n)$. The \scshape Construct $\hat{H}$ Algorithm \normalfont is useful in the following sense. Suppose $i \in I$ is a check node adjacent to several variable nodes $j \in J$ such that $x^*_j$ is non-integral. If $\hat{H}$ has such a row $i$ then we use Proposition \ref{prop1} and Proposition \ref{prop2} to construct Forbidden Set Inequalities which cut off the fractional optimal solution. Specifically we construct the inequalities (\ref{ineq:validcutsodd}) or (\ref{ineq:validcutseven}). We refer to these inequalities as new Forbidden Set Inequalities. Note that $N_i$ in the original $H$ matrix and $\hat{N}_i$ in $\hat{H}$ are different index sets. First we calculate  
\begin{align}
k_i=\left|\{ h \in \hat{N}_i | x^*_h=1 \}\right|.
\end{align}
If $k_i$ is odd we use the inequality
\begin{align}
\sum_{h\in \hat{N}_i:x_h^{*}=1}x_h - x_j - \sum_{h\in \hat{N}_i:x_h^{*}=0}x_h \leq k_i -1, \label{ineq:validcutsodd}
\end{align}
otherwise, $k_i$ is even, i.e.
\begin{align}
\sum_{h\in \hat{N}_i:x_h^{*}=1}x_h + x_j - \sum_{h\in \hat{N}_i:x_h^{*}=0}x_h \leq k_i. \label{ineq:validcutseven}
\end{align}
\begin{thm} \label{thm:newforbidset}
Let $(x^*,z^*) \in \mathbb{R}^n \times \mathbb{R}^m$ be the optimal solution of the current RIPD formulation such that $x^*$ is non-integral. If there exists a $\hat{H}_{i,.}$ such that $\hat{H}_{i,j}=1$ and $x^*_j$ is non-integral for exactly one $j \in J$ then the new forbidden set inequality is a valid inequality which is violated by $x^*$. \label{theo:validcut}
\end{thm}

\noindent{\underline{Proof:}} We have to show that:
\begin{enumerate}
	\item For $k_i$ odd $\left[\text{even}\right]$ the inequality (\ref{ineq:validcutsodd}) $\left[(\ref{ineq:validcutseven})\right]$ is violated by $x^*$.
	\item For $k_i$ odd $\left[\text{even}\right]$ the inequality (\ref{ineq:validcutsodd}) $\left[(\ref{ineq:validcutseven})\right]$ is satisfied for all $x \in C$.
\end{enumerate}

Let $i \in I$ be a row of the reconstructed matrix $\hat{H}$. We obtain $i$ by performing elementary row operations in $GF(2)$ on the rows of the original $H$ matrix. Therefore it holds that $\hat{H}_{i,.}x=0\text{ mod}2$ for all $x \in C$. We show the proof for $k_i$ odd. When $k_i$ is even the proof is analogous. 

1) Let $k_i$ be an odd number. For $x^*$, since $0<x_j^*<1$ the left hand side of (\ref{ineq:validcutsodd}) is larger than the right hand side thus $x^*$ violates (\ref{ineq:validcutsodd}).\\

2)Suppose $k_i$ is odd and $x^*$ is the optimal solution of RIPD. Our aim is to show that (\ref{ineq:validcutsodd}) is satisfied by all codewords $x \in C$. First we define 
\begin{align*}
\delta_i(x)=\sum_{j \in \hat{N}_i}x_j.
\end{align*}

Next we rewrite (\ref{ineq:validcutsodd}) as 
\begin{align}
\sum_{j \in \hat{N}_i}a_jx_j \leq k_i -1 \text{ where } a_j \in \{-1,1\}. \label{ineq:checkconstraint}
\end{align}

We also define the index sets
\begin{align*}
&S^+ = \{j \in \hat{N}_i : a_j=1\} \text{ with } \left|S^+\right|=k_i.\\
&S^- = \{j \in \hat{N}_i : a_j=-1\} \text{ with } \left|S^-\right|= \left|\hat{N}_i\right|-k_i.
\end{align*} 

\underline{Case 1} For any $x \in C$ it holds that $\delta_i(x) \leq k_i-1$:
\begin{align*}
\sum_{j \in \hat{N}_i}a_jx_j \leq k_i-1 \text{ is fulfilled.}
\end{align*}  

\underline{Case 2a} For any $x \in C$ it holds that $\delta_i(x) \geq k_i+1$: At most $k_i$ of indices $j \in \hat{N}_i$ where $x_j=1$ can be in $S^+$. Thus there is at least one index $j \in \hat{N}_i$ with $x_j=1$ in $S^-$. Consequently 
\begin{align*}
\sum_{j \in \hat{N}_i}a_jx_j \leq k_i-1.
\end{align*}

\underline{Case 2b} For any $x \in C$ it holds that $\delta_i(x) = k_i$: If there is at least one index $j \in S^-$ with $x_j=1$ then
\begin{align*}
\sum_{j \in \hat{N}_i}a_jx_j \leq k_i-1.
\end{align*}
Otherwise all $j \in \hat{N}_i$ with $x_j=1$ are in $S^+$. Then for row $i$, $\hat{H}_{i,.}x=1\text{ mod}2$ since $k_i$ is odd and therefore the contradiction $x \notin C$. \qeda

Note that it is possible that each row of $\hat{H}$ has at least two $j \in J$ such that $\hat{H}_{i,j}=1$ and $x^*_j$ is non-integral. In this case no new forbidden set inequality can be found using \scshape Cut Generation Algorithm \normalfont $2$.\\ 

\begin{tabular}{|p{0.9\columnwidth}|} 
\hline \\
\textbf {\underline{\scshape Cut Generation Algorithm \normalfont $2$}}\\\\
\textbf {Input} :  Optimum of RIPD s.t. $x^*$ non-integral, $\hat{H}$. \\
\textbf {Output} : New forbidden set inequality or error.\\
\textbf {1} : Set $i=1$.\\
\textbf {2} : If there is exactly one $j \in J$ such that $\hat{H}_{i,j}=1$ and $x^*_j \in (0,1)$, then calculate $k_i$ and go to $3$. Else go to $4$.\\
\textbf {3} : If $k_i$ is odd $\left[\text{even}\right]$ construct (\ref{ineq:validcutsodd}) $\left[(\ref{ineq:validcutseven})\right]$. Terminate. \\
\textbf {4} : Set $i=i+1$. If $i \leq m$ then go to $2$. Else output {\tt error}.\\
\hline
\end{tabular}\\[0.1cm]

The complexity of \scshape Cut Generation Algorithm \normalfont $2$ is in $O(mn)$ since in the worst case each entry of $\hat{H}$ has to be visited once . 

We are now able to formulate our separation algorithm. In the first iteration, $x^*$ can be found by hard decision decoding. In all of the following iterations RIPD does not necessarily have an optimal solution with integral $x^*$. If the vector $(x^*,z^*)$ is integral then the optimal solution to $IPD$ is found. If $x^*$ is integral but $z^*$ is non-integral we apply \scshape Cut Generation Algorithm \normalfont $1$ to construct Forbidden Set Inequalities. Although adding any forbidden set inequality suffices to cut off the non-integral solution $(x^*,z^*)$ we add all Forbidden Set Inequalities induced by all non-integral $z_i$ based on the thought that they may be useful in future iterations. If $x^*$ is non-integral we first employ the \scshape Construct $\hat{H}$ Algorithm \normalfont. Then we check in \scshape Cut Generation Algorithm \normalfont $2$ if there exists a row $\hat{H}_{i,.}$ such that there exists exactly one $j \in J$ where $\hat{H}_{i,j}=1$ and $x^*_j$ is non-integral. If such a row does not exist, then the \scshape Cut Generation Algorithm \normalfont $2$  outputs an error. Otherwise we know from Theorem \ref{thm:newforbidset} that there exists a new forbidden set inequality which cuts off $x^*$. In $\hat{H}$ there may exist several rows from which we can derive new Forbidden Set Inequalities. In this case we add all new Forbidden Set Inequalities to the formulation RIPD with the same reasoning as before. The \scshape New Separation Algorithm \normalfont stops if either $(x^*,z^*)$ is integral which leads to an ML Codeword or \scshape Cut Generation Algorithm \normalfont $2$  returns an {\tt error} which means no further cuts can be found.

\begin{tabular}{|p{0.9\columnwidth}|} 
\hline \\
\textbf {\underline{\scshape New Separation Algorithm \normalfont}} .\\\\
\textbf {Input} :  Cost vector $c$, matrix $H$.  \\
\textbf {Output} : Current optimal solution $x^*$.\\
\textbf {1} : Solve RIPD.\\
\textbf {2} : If the optimal solution $(x^*,z^*)$ is integral then go to $6$. Otherwise go to $3$.\\
\textbf {3} : If $x^*$ is integral, then call \scshape Cut Generation Algorithm \normalfont $1$. Add the constraints to formulation RIPD, go to $1$. If $x^*$ is non-integral go to $4$.\\
\textbf {4} : Call \scshape Construct $\hat{H}$ Algorithm \normalfont. Go to 5.\\
\textbf {5} : Call \scshape Cut Generation Algorithm \normalfont $2$. If the output is {\tt error} then go to $6$. Otherwise add the new constraint to formulation RIPD, go to $1$.\\   
\textbf {6} : Output $x^*$ and terminate.\\
\hline
\end{tabular}\\[0.1cm]

Two strategies which may be used in the implementation of the \scshape New Separation Algorithm \normalfont are:
\begin{enumerate}
	\item Add all valid cuts which can be obtained in one iteration.
	\item Add only one of the valid cuts which can be obtained in one iteration.
\end{enumerate}

There is a trade-off between Strategies $1$ and $2$, since strategy $1$ means less iterations with large LP problems and Strategy $2$ means more iterations with smaller LP problems. We empirically tested Strategies $1$ and $2$ on the three codes described in the follwing section. For all the three codes Strategy $1$ outperformed Strategy $2$ in terms of running time and decoding success. 

\section{Numerical Results}
We compare the communication performance of our separation algorithm with the standard LP decoding \cite{FeWaKa}, BP decoding, and the reference curve resulting from ML decoding. The latter results from modeling and solving IPD using CPLEX 9.120 \cite{cplex} as the IP solver. These four algorithms, LP decoding (by Feldman et al. or Taghavi et al.), BP, \scshape New Separation Algorithm \normalfont, and ML Decoding(IP, CPLEX) are tested on two LDPC (one regular and one irregular) and one BCH code considering transmission over Additive White Gaussian Noise (AWGN) channels. Additionally we present for our separation algorithm the min, max and average values for the number of iterations, the number of generated Gomory cuts and the number of generated RPC cuts in tables I, II, III. We selected the $(64,32)$ irregular LDPC code, Tanner's $(155, 64)$ group structured LDPC code \cite{Tanner} and the $(63,39)$ BCH code for our tests. The first LDPC code is constructed with Progressive Edge Growth algorithm. Tanner's $(155, 64)$ LDPC code, which has minimum distance of 20 and girth of 8, is constructed as described in \cite{Tanner}. The Frame Error Rate (FER) against signal to noise ratio (SNR) measured in $E_S/N_0$ is shown in Figures \ref{fig:64-32} to \ref{fig:63-39}. We used $200$ iterations for BP decoding of $(64,32)$ irregular LDPC and Tanner's $(155, 64)$ LDPC code.  

\reffig{fig:64-32} shows the results for the irregular $(64,32)$ LDPC code with degree distribution \footnote{Irregular LDPC codes are described by variable node degree distribution $f_{i}$ and check node degree distribution $g_{i}$, where $f_{i}$ and $g_{i}$ represents the fraction of variable nodes and check nodes with degree $i$ respectively.} $f_{[2,3,5,6]}=[f_2=\frac{1}{2},f_3=\frac{1}{4},f_5=\frac{1}{8},f_6=\frac{1}{8}]$, $g_{[6]}=[1]$. Our separation algorithm performs by roughly $0.5dB$ better than LP decoding for this LDPC code. It is important to note that the communication performance of the \scshape New Separation Algorithm \normalfont is superior to the BP algorithm here.

The results for the Tannner's $(155,64)$ LDPC code are plotted in \reffig{fig:155-93}. Performance of the BP and standard LP decoding is very similar in this case whereas the \scshape New Separation Algorithm \normalfont gains around $0.4dB$ compared to both. It is worthwhile mentioning that BP decoding and our separation algorithm have a performance degradation of $>0.8dB$ compared to ML decoding for this group structured LDPC code.

LP decoding via Forbidden Set Inequalities introduced in \cite{FeWaKa} cannot be used for high density codes since the number of constraints is exponential in the check node degree. This causes a prohibitive usage of memory in the phase of building the LP model. The adaptive approach of \cite{TagSie} overcomes this shortcoming and yet performs as good as LP decoding (see Section III). Therefore we used this method in the comparison of algorithms when decoding a dense (63,39) BCH code. The results for this code are shown in \reffig{fig:63-39}. It should also be noted that BP decoding does not work for this type of codes due to the dense structure of their parity check matrix. Our approach is one of the first attempts (see \cite{DrYeYi}) to decode dense codes using mathematical programming approaches. Although the gap between ML decoding and our separation algorithm increases to roughly $1dB$, the results obtained by our algorithm are substantially better (more than $2dB$) than the results obtained by adaptive LP decoding.

To summarize, our separation algorithm improves LP decoding significantly for all three test setups. This improvement is due to new Forbidden Set Inequalities found by \scshape Cut Generation Algorithm \normalfont $2$. The constraints added by this algorithm are based on the rows of the alternative representations of the $H$ matrix. These rows can also be interpreted as redundant parity checks. Consequently, the family $\Lambda$ of inequalities we use includes a subset of the Forbidden Set Inequalities which can be derived from redundant parity checks and $\Lambda$ is larger than the original family of Forbidden Set Inequalities.

Regarding the complexity of the \scshape New Separation Algorithm \normalfont, we present the minimum, average, and maximum number of iterations, cuts introduced by the \scshape Cut Generation Algorithm $1$ \normalfont (shown in Gomory cuts column) and the number of cuts introduced by the \scshape Cut Generation Algorithm \normalfont $2$ (shown in RPC cuts column) in the tables \ref{tab:1}, \ref{tab:2}, and \ref{tab:3} for the codes $(64,32)$, $(155,64)$, and $(63,39)$ respectively. Note that the number of iterations can be considered as the number of times we call the LP solver.
\begin{figure}[htp!]
\centering 
\includegraphics[width=1.0\columnwidth, keepaspectratio]{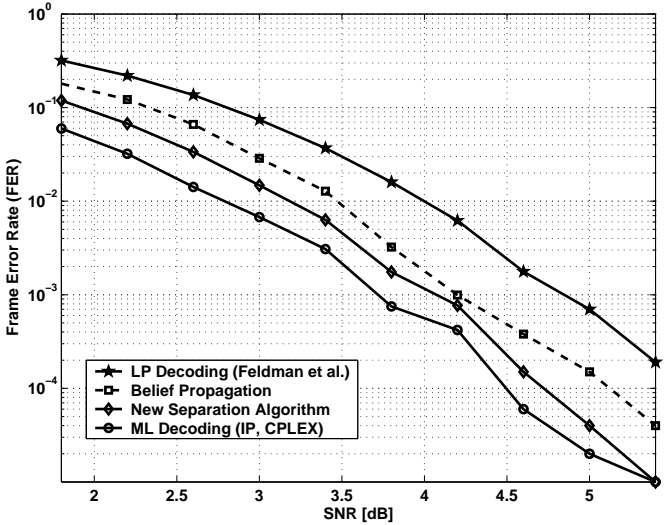}
\vspace{-0mm}
\caption{Decoding performance of an irregular LDPC code (64,32).} 
\label{fig:64-32}
\vspace{-0mm}
\end{figure}

\begin{table*}[htp!]
\centering
\begin{tabular}{r@{.}l||c|r@{.}l|c||c|r@{.}l|c||c|r@{.}l|c}
\multicolumn{2}{c||}{}& \multicolumn{4}{c||}{\textbf{Number of LPs solved}} & \multicolumn{4}{c||}{\textbf{Number of Gomory cuts}} & \multicolumn{4}{c}{\textbf{Number of RPC cuts}} \\ \hline

\multicolumn{2}{c||}{\textbf{SNR}}&		\textbf{Min} & \multicolumn{2}{c|}{\textbf{Average}} & \textbf{Max} &		\textbf{Min} & \multicolumn{2}{c|}{\textbf{Average}} & \textbf{Max} &		\textbf{Min} & \multicolumn{2}{c|}{\textbf{Average}} & \textbf{Max}  \\ \hline
1&8&	2&	5&942&	20&	2&	21&296&	42&	0&	33&619&	207 \\
2&2&	1&	4&896&	21&	0&	19&187&	41&	0&	21&465&	227 \\
2&6&	1&	4&196&	19&	0&	17&569&	42&	0&	13&138&	177 \\
3&0&	1&	3&48&	  16&	0&	15&07&	40&	0&	6&895&	180 \\
3&4&	1&	3&005&	19&	0&	13&228&	39&	0&	2&917&	145 \\
3&8&	1&	2&725&	12&	0&	11&254&	36&	0&	1&513&	119 \\
4&2&	1&	2&446&	11&	0&	9&738&	31&	0&	0&428&	111 \\
4&6&	1&	2&297&	10&	0&	8&195&	32&	0&	0&27&	52 \\
5&0&	1&	2&134&	6&	0&	7&055&	31&	0&	0&079&	25 \\
5&4&	1&	1&977&	6&	0&	5&585&	23&	0&	0&014&	6 \\
5&8&	1&	1&872&	6&	0&	4&448&	18&	0&	0&012&	12 \\
\end{tabular}
\caption{Iterations and cuts derived for (64,32) LDPC code.}
\label{tab:1}

\end{table*}

\begin{figure}[htp!]
\centering 
\includegraphics[width=1.0\columnwidth, keepaspectratio]{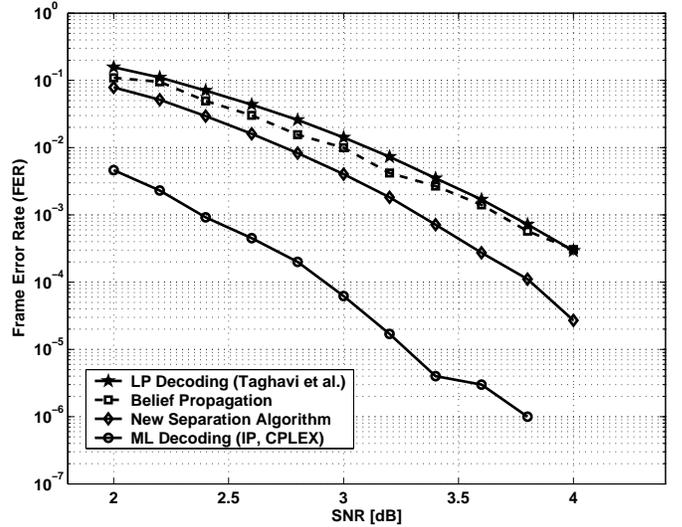}
\vspace{-0mm}
\caption{Decoding performance of Tanner's $(155, 64)$ LDPC code.} 
\label{fig:155-93}
\vspace{-0mm}
\end{figure}

\begin{table*}[htp!]
\centering
\begin{tabular}{r@{.}l||c|r@{.}l|c||c|r@{.}l|c||c|r@{.}l|c}
\multicolumn{2}{c||}{}& \multicolumn{4}{c||}{\textbf{Number of LPs solved}} & \multicolumn{4}{c||}{\textbf{Number of Gomory cuts}} & \multicolumn{4}{c}{\textbf{Number of RPC cuts}} \\ \hline

\multicolumn{2}{c||}{\textbf{SNR}}&		\textbf{Min} & \multicolumn{2}{c|}{\textbf{Average}} & \textbf{Max} &		\textbf{Min} & \multicolumn{2}{c|}{\textbf{Average}} & \textbf{Max} &		\textbf{Min} & \multicolumn{2}{c|}{\textbf{Average}} & \textbf{Max}  \\ \hline
2&0&	2&	6&093 &	20 &	20&	60&235&	94&	 0&	74&161&	594 \\
2&2&	2&	5&343 &	22 &	19&	57&148&	100& 0&	48&667&	595 \\
2&4&	2&	4&828 &	21 &	19&	54&013&	94&	 0&	31&713&	640 \\
2&6&	2&	4&363 &	23 &	14&	50&817&	92&	 0&	20&254&	549 \\
2&8&	2&	3&954 &	18 &	15&	47&265&	96&	 0&	12&65&	468 \\
3&0&	2&	3&798 &	26 &	16&	45&324&	98&	 0&	10&776&	632 \\
3&2&	2&	3&47  &	17 &	16&	42&2&	  79&	 0&	4&211&	431 \\
3&4&	2&	3&158 &	19 &	11&	38&381&	81&	 0&	1&293&	508 \\
3&6&	2&	3&13  &	13 &	6&	36&478&	76&	 0&	1&122&	228 \\
3&8&	2&	2&911 &	10 &	3&	34&085&	76&	 0&	0&324&	252 \\
4&0&	2&	2&81  &	12 &	7&	31&529&	66&	 0&	0&298&	238 \\
4&2&	2&	2&725 &	 9 &	7&	29&576&	68&	 0&	0&146&	78 \\
\end{tabular}
\caption{Iterations and cuts derived for $(155,64)$ Tanner code.}
\label{tab:2}

\end{table*}

\begin{figure}[htp!]
\centering 
\includegraphics[width=1.0\columnwidth, keepaspectratio]{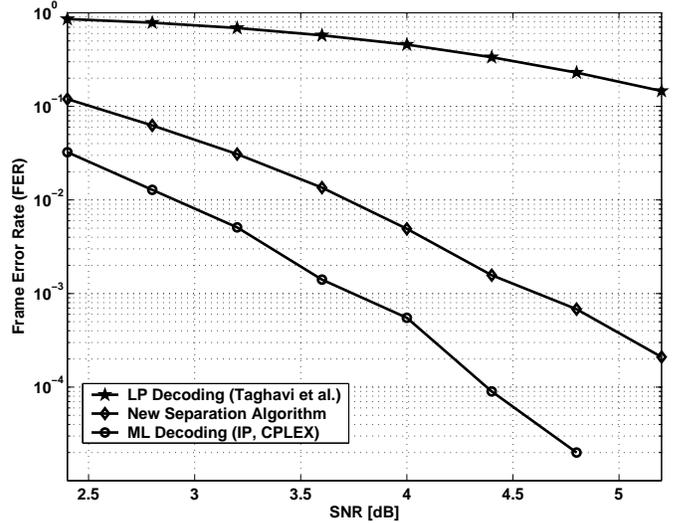}
\vspace{-0mm}
\caption{Decoding performance of a BCH code (63,39).} 
\label{fig:63-39}
\vspace{-0mm}
\end{figure}

\begin{table*}[htp!]
\centering
\begin{tabular}{r@{.}l||c|r@{.}l|c||c|r@{.}l|c||c|r@{.}l|c}
\multicolumn{2}{c||}{}& \multicolumn{4}{c||}{\textbf{Number of LPs solved}} & \multicolumn{4}{c||}{\textbf{Number of Gomory cuts}} & \multicolumn{4}{c}{\textbf{Number of RPC cuts}} \\ \hline

\multicolumn{2}{c||}{\textbf{SNR}}&		\textbf{Min} & \multicolumn{2}{c|}{\textbf{Average}} & \textbf{Max} &		\textbf{Min} & \multicolumn{2}{c|}{\textbf{Average}} & \textbf{Max} &		\textbf{Min} & \multicolumn{2}{c|}{\textbf{Average}} & \textbf{Max}  \\ \hline
2&4&	1&	10&186&	24&	0&	24&993&	56&	0&	64&173&	200 \\
2&8&	1&	8&802&	21&	0&	23&464&	57&	0&	50&382&	175 \\
3&2&	1&	7&649&	22&	0&	22&083&	53&	0&	39&76&	180 \\
3&6&	1&	5&911&	22&	0&	19&401&	63&	0&	25&184&	175 \\
4&0&	1&	4&967&	21&	0&	17&743&	54&	0&	17&729&	179 \\
4&4&	1&	4&111&	20&	0&	15&379&	60&	0&	11&612&	176 \\
4&8&	1&	3&249&	18&	0&	12&941&	59&	0&	6&508&	177 \\
5&2&	1&	2&703&	18&	0&	10&944&	43&	0&	4&002&	143 \\
\end{tabular}
\caption{Iterations and cuts derived for (63,39) BCH code.}
\label{tab:3}

\end{table*}

\section{Conclusion}
In this paper we proposed a new IP formulation and its LP relaxation. Instead of solving the optimization problem, we solve the separation problem. The indicator variables $z$ yield an immediate recognition of parity violations and efficient generation of cuts. We used on one hand the Forbidden Set Inequalities of \cite{FeWaKa} which are a subset of all possible Gomory cuts. On the other hand we showed how to generate efficiently new cuts based on redundant parity checks. Note that the rows in our $\hat H$ matrix can be considered as redundant parity checks. It is known that RPC cuts improve the LP decoding via tightening the fundamental polytope \cite{FeWaKa}, \cite{TagSie}. However RPC generating approaches known to us cannot verify if the particular RPC really introduces a cut or not. Another open question addresses the configuration $S$ to be used for the RPC. In our approach, once we ensure that there is only one $j \in N_{i}$ with non-integral $x^*_j$ in row $\hat{H}_{i,.}$, we can immediately find the configuration $S$ and thus the new forbidden set inequality (\ref{ineq:validcutsodd}) or (\ref{ineq:validcutseven}). Additionally, Theorem \ref{theo:validcut} states that the new forbidden set inequality is a valid inequality which cuts off the fractional optimal solution $(x^*,z^*)$.

These theoretical improvements are supported with empirical evidence. Compared to state of the art (adaptive) LP decoding our algorithm is superior in terms of frame error rate for all the codes we have tested. Moreover, it is competitive to the results obtained by BP decoding. In contrast to the latter, our approach is applicable to codes with dense parity-check matrix and offers a possibility to decode such codes.

One future research direction is to find new cut families when \scshape Cut Generation Algorithm \normalfont $2$ stops. The polyhedral structure of the ML decoding will be further investigated. This will yield a branch-and-cut algorithm which we expect to further extend the applicability of our approach.


%

\appendices


\section*{Acknowledgment}
We would like to thank Pascal O. Vontobel for his constructive comments \& suggestions and our colleague Daniel Schmidt for his initial work related to IPD formulation presented in this paper.  We gratefully acknowledge partial financial support by the Center of Mathematical and Computational Modeling of the University of Kaiserslautern. 

\ifCLASSOPTIONcaptionsoff
  \newpage
\fi

\end{document}